# A method for determination of gamma-ray direction in space

Serkan Akkoyun

*Cumhuriyet University, Faculty of Science, Department of Physics, Sivas-Turkey*

**Abstract**

Gamma-ray bursts (GRBs) are short and most intense bursts of gamma-rays that come from random direction in space. Their origin are still unknown and they originate likely from cosmological distances, probably after birth of a new black hole or death of a giant star. In this work, Geant simulations of a detector array whose aim is to identify gamma-ray directions in space were performed and a method for this identification was developed. The array consists of three quadratic NaI(Tl) scintillators which are facing different directions and the method is based on the difference of the counts registered in these three detectors. By using the method the gamma-ray directions are obtained with $10^o$ accuracy. This form of the array which can scan three dimensions in space is crucial to pinpoint origin of the GRBs. The array would also be applicable in various fields where identifications of the gamma-ray directions are necessary.

Corresponding author e-mail: sakkoyun@cumhuriyet.edu.tr, Tel: +903462191010x1413

## 1. Introduction

Gamma-ray bursts (GRBs) are short and intense flashes of photons that come from random direction in space and last from fraction of a second up to few minutes [1]. The origin of the GRBs are still unknown and they originate from cosmological distances, probably after birth of a new black hole or death of a giant star (hyper/super-nova) [2]. These bursts emitting in the energy range between about 0.1 MeV to few MeV often outshine other gamma-ray sources. These short-lived bursts are usually followed by a much longer-lived signal. This so-called "afterglow" [3-5] emitted at longer wavelengths allows us to pinpoint the origin of the gamma-ray burst that is not possible from the short-lived gamma-ray signal alone. The sources of most GRBs are billions of light years away from Earth, indicating that the explosions are both extremely energetic and rare [6]. The topic of GRBs has been reviewed over the years in several works [1, 7-11].

GRBs remain one of the most interesting objects in high energy astrophysics almost 50 years after their discovery. A new era in GRB studies have been opened about 20 years ago with the launch of the Compton Gamma-Ray Observatory (CGRO) [1]. The most significant results came from the all-sky observing by the Burst and Transient Experiment (BATSE) on CGRO complemented by data from the OSSE [12], Comptel [13] and EGRET [14] experiments. The next major discovery was made by satellite BeppoSAX, which was able to produce the first quick and small error boxes of GRBs' positions in the sky [15]. This allows observing position of GRB several hours or days after their explosion. Nowadays the detections of GRBs are performed by HETE, Integral, Swift and AGILE satellites [16].

Several techniques have been developed in order to obtain directional information of gamma-rays from GRBs or the other sources [17-22]. Accurate burst locations are necessary for solving GRB mystery. In this work, new types of a detector array have been designed to obtain directional information of incident gamma-rays from mainly GRBs. This array consists of three quadratic NaI(Tl) detectors which are perpendicular to each other viewing different directions in space. By making use of three energy spectra obtained in these three detectors, determination of gamma-ray direction can be achieved. This array can scan three dimensions in space and identify the location of the bursts within about $10^{o}$ accuracy. The detector array would be applicable in various fields such as in measurements of terrestrial gamma-rays [23], in determination of the location of missing gamma-emitting radioactive sources, in identification of radioactive areas and in explanation dose rate differences found by car-borne measurement.

## 2. Experimental setup and method

The detector array designed for placing on a satellite is consist of three quadratic NaI(Tl) crystals which are perpendicular each other, as shown in Fig.1. In ideal case, the thickness of the detector should be zero in order to perfectly determine gamma-ray directions. Namely as can be seen in Fig.1.a, in order to obtain directional information of the gamma-rays coming from the source along the z direction, they must hit the whole surface of the detector A and must not hit the detector B and C. In this case, one can say that gamma-rays are measured only by the detector A. But at this time, the detector A will not be able to detect gamma-rays due to its extremely thinness. After several investigations by considering both photopeak efficiencies of each individual detector and efficiency of the array, the dimensions of the detectors were chosen as $5 \times 5 \times 1.2\, cm^3$. Then the total weight of the crystals is 330 g and the estimated total weight of the detectors including electronic and mechanical parts is 600 g. Then after given an optimal thickness to the detectors, in the case of Fig.1.a, the majority of the gamma-ray counts are registered in the detector A. Still, accurately determinations of the gamma-ray directions are possible. Already, the method developed in this work is based on the relative counts belonging to the different detectors.

*2.1 Simulations*

The simulation of the detectors and gamma-ray interactions were performed by Geant simulation program [24]. The detector array mentioned above located in a vacuum environment (denoted by big cubes with dot-dashed lines in Fig.1) which represents the space. Due to the big distance between detectors and gamma-ray source, the gamma-ray beams reached to the detectors are considered to be in parallel. It is also assumed that the detectors can measure gamma-rays only by their front and lateral faces due to the electrical and mechanical parts of the detectors are placed on the shielded backside of the detectors.

## 2.2 Method

Changing incoming directions of the gamma-ray beams changes the each photopeak counts recorded in three detectors which view different directions in space. The relative responses of the detectors are used for the determination of the gamma-ray directions. Because of the rotational motion of the satellite around the world and itself, it is essential to know measurement moment of the gamma-rays for getting true gamma-ray directions. This rotational motion of the satellite allows scanning entire space in three dimensions by the detector array. According to the method, as an illustration in Fig.1.a, because of the surface area of the detector A is bigger than that of detector B and C viewing by incoming gamma-rays, more gamma-rays enter the detector A. Therefore, the photopeak counts belonging to the detector A is expected more than B and C. Eventually in this case, one can predict the incoming direction of the gamma-rays towards the surface of detector A. Similarly in Fig.1.b and 1.c, it is easily said that the gamma-ray directions are towards the surface of detector B and C respectively.

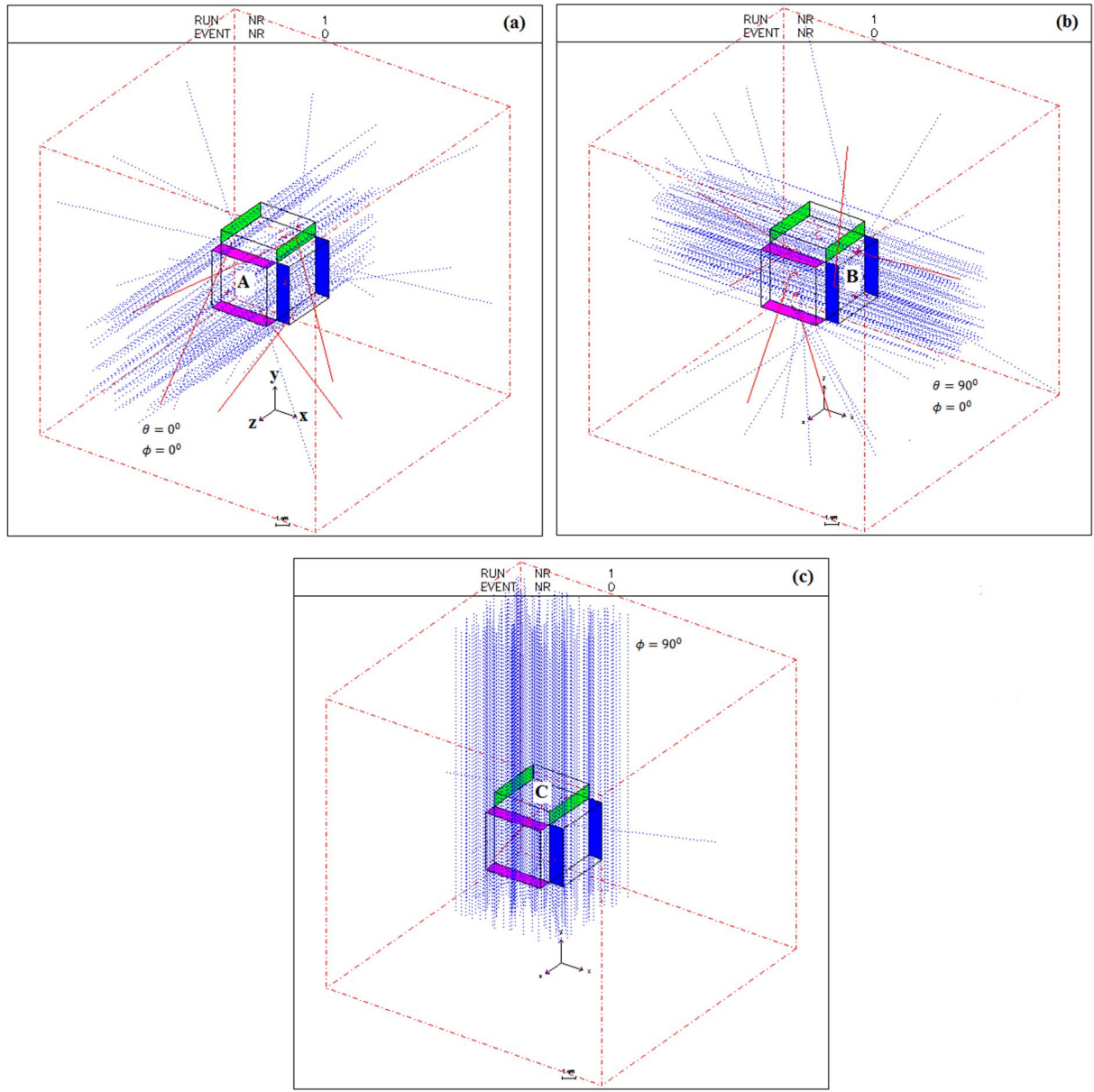

**Fig.1** Experimental setup with three NaI crystals indicated in different colours. The gamma-rays enter perpendicularly the detector surfaces A (a), B (b) and C (c).

## 3. Application of the method

In order to determine gamma-ray directions, the photopeak counts in three different detectors were obtained. The photopeak counts registered in detectors A, B and C were indicated by A, B and C. As T represents the total photopeak counts registered in all three detectors, the count ratios A/T, B/T and C/T can be useful for identification of the gamma-ray directions. In this study, the systematics of the angle of incident gamma-ray can be explained by helping Fig.1s. The $\theta$ and $\phi$ angles which are perpendicular to the surface of the detector A are $0^o$ and $0^o$ respectively. By increasing $\theta$, the gamma-ray beam direction rotates counter-clockwise about the y-axis and after rotation of $90^o$ finally becomes perpendicular to the surface of the detector B. In this situation the final angles are $\theta = 90^o$ and $\phi = 0^o$. Besides, the counter-clockwise rotation about the z-axis increases the φ angle. By increasing φ, the gamma-ray beam direction which were perpendicular to the surface of the detector B ($\phi = 0^o$) becomes perpendicular to surface of the detector C after rotation of $90^o$.

The analyses were performed by changing one of the angle by $30^o$ step while the other is constant. The result were given in Table.1 and 2 for 1,3 and 10 MeV gamma-rays with constant $\theta$ and $\phi$. As can be clearly seen in the Table.1, at $\theta = 90^o$ and $\phi = 0^o$ the detector B and C have their maximum and minimum counts ratios respectively, as expected. By increasing $\phi$ value, the ratio of the detector B decreases while the detector C's increases and reaches their minimum and maximum at $\theta = 90^o$ and $\phi = 90^o$ for the detector B and C, respectively. This situation seems only for 1 and 3 MeV energy values. But there is an opposite situation for 10 MeV gamma-rays. Due to their long range in this high energy, the thickness of the detector is not enough to efficiently detect them. Besides, the incoming gamma-rays enter the detector A and C by lateral sides and are detected more probably although their surfaces viewed by gamma-rays are smaller. So, while the count ratio belonging to the detector C decreases, the detector B's increases. Furthermore, effect of changing $\phi$ value do not appreciably affect the ratio belonging to the detector A due to the gamma-rays always enter the detector in its lateral side.

**Table.1** Changing of photopeak counts ratios by $\phi$ for fixed $\theta$. Analyses were performed for 1, 3 and 10 MeV gamma-rays.

| $E_\gamma (MeV)$ | $\theta = 90^o$ Angle | Photopeak count ratios | | |
|---|---|---|---|---|
| | | A/T | B/T | C/T |
| 1 | $\phi = 0^o$ | 0.29 | 0.41 | 0.30 |
| | $\phi = 30^o$ | 0.28 | 0.37 | 0.35 |
| | $\phi = 60^o$ | 0.27 | 0.36 | 0.37 |
| | $\phi = 90^o$ | 0.30 | 0.30 | 0.40 |
| 3 | $\phi = 0^o$ | 0.31 | 0.37 | 0.32 |
| | $\phi = 30^o$ | 0.30 | 0.36 | 0.34 |
| | $\phi = 60^o$ | 0.31 | 0.34 | 0.35 |
| | $\phi = 90^o$ | 0.31 | 0.32 | 0.37 |
| 10 | $\phi = 0^o$ | 0.35 | 0.31 | 0.34 |
| | $\phi = 30^o$ | 0.34 | 0.35 | 0.31 |
| | $\phi = 60^o$ | 0.35 | 0.35 | 0.30 |
| | $\phi = 90^o$ | 0.33 | 0.37 | 0.30 |

Similar effect can be seen for effect of changing $\theta$ value in Table.2. Maximum count ratio belonging to the detector A and minimum count ratio belonging to the detector B is at $\phi = 0^o$ and $\theta = 0^o$ as expected. These ratios increase for the detector A and decrease for the detector B with increasing $\theta$. Finally the ratio belonging to the detector A reaches its minimum at $\phi = 0^o$ and $\theta = 90^o$ at which the ratio belonging to the detector B is in its maximum. This situation seems only for 1 and 3 MeV energy values and the ratio belonging to the detector C remain almost constant as mentioned above. Thereby in this method, the gamma-ray direction can be determined by using knowledge of the count ratios of each three detector.

**Table.2** Changing of photopeak counts ratios by $\theta$ for fixed $\phi$. Analyses were performed for 1, 3 and 10 MeV gamma-rays.

| $E_\gamma (MeV)$ | $\phi = 0^o$ Angle | Photopeak count ratios | | |
|---|---|---|---|---|
| | | A/T | B/T | C/T |
| 1 | $\theta = 0^o$ | 0.41 | 0.30 | 0.29 |
| | $\theta = 30^o$ | 0.37 | 0.35 | 0.28 |
| | $\theta = 60^o$ | 0.35 | 0.37 | 0.28 |
| | $\theta = 90^o$ | 0.29 | 0.41 | 0.30 |
| 3 | $\theta = 0^o$ | 0.36 | 0.31 | 0.33 |
| | $\theta = 30^o$ | 0.36 | 0.33 | 0.31 |
| | $\theta = 60^o$ | 0.33 | 0.36 | 0.31 |
| | $\theta = 90^o$ | 0.31 | 0.37 | 0.32 |
| 10 | $\theta = 0^o$ | 0.25 | 0.36 | 0.39 |
| | $\theta = 30^o$ | 0.29 | 0.36 | 0.35 |
| | $\theta = 60^o$ | 0.37 | 0.28 | 0.35 |
| | $\theta = 90^o$ | 0,35 | 0.31 | 0.34 |

The results are also given for fixed $\phi$ values and different gamma-ray energies in Figs.2. In Figs.2.a and 2.b, for 1 and 3 MeV gamma-rays and fixed $\phi$, decreases and increases in count ratios of the detector A and B, respectively, are obviously seen while for the detector C's stays nearly same level. By increasing $\theta$, the gamma-ray beams shift toward the detector B from A. Therefore, the surface area of the detector A viewed by gamma-rays decreases and the area of the detector B increases. Additionally it is evidently noticed that, the count ratios of A and B have same values around the 45° and are symmetric on both sides of that point. Moreover, at $\phi = 45^o$ and $\theta = 45^o$, all three detector count ratios are equal due to the fact that exact symmetry of the array according to the gamma-ray beams. In Fig.2.c, for 10 MeV high-energy gamma-rays, an opposite situation appears. Accordingly, the ratio of the detector A increases while the detector B's decreases by increasing $\theta$, as mentioned before.

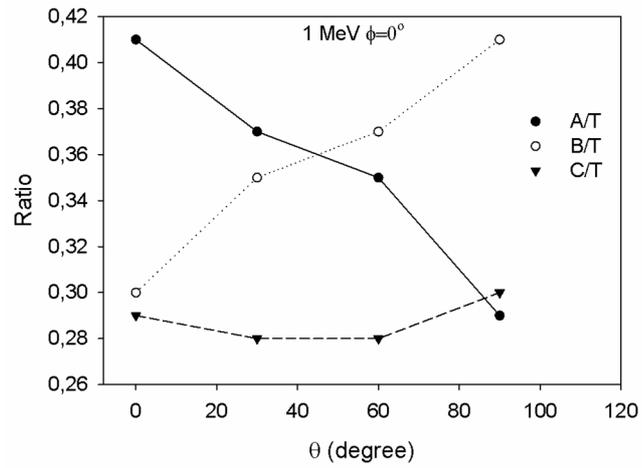

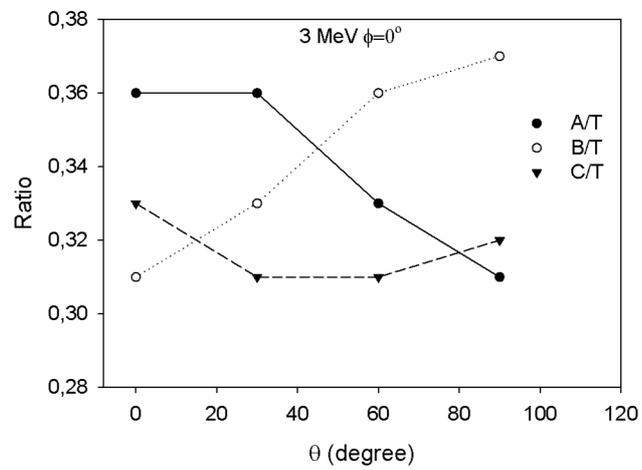

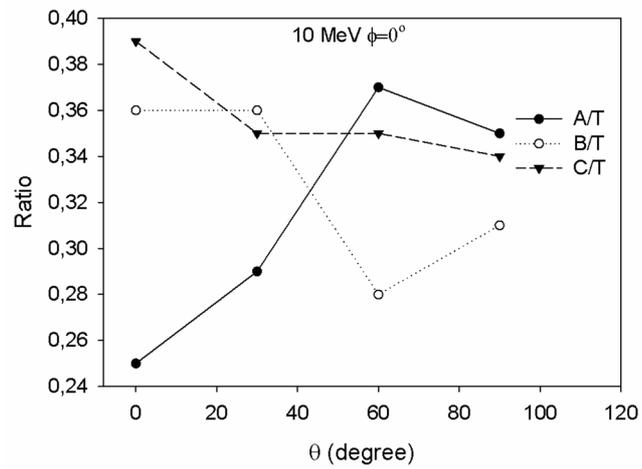

**Fig.2** The changing of detector photopeak count ratios by θ for 1 MeV (a), 3 MeV (b) and 10 MeV (c) gamma-rays.

## 4. Results

The plots which can help for determination of the gamma-ray directions were given in Figs.3. A/T, B/T and C/T count ratios versus $\theta$ were shown for different $\phi$ values in Figs.3a, 3b and 3c respectively. As is seen in the Fig.3.a, the ratios decrease by increasing $\theta$ for $\phi = 0^o, 30^o$ and $60^o$ because of shifting gamma-ray beam from surface of the detector A to B. So, the ratios related to detector B increase for same angles (Fig.3.b). There is a different situation for $\phi = 90^o$. In this case, change of the $\theta$ does not alter the ratios, due to the gamma-ray beam stays perpendicular to lateral surface of the detector A and B in every $\theta$ values. From the Fig.3.c, one can see that the count ratios for every $\phi$ values stay nearly steady even $\theta$ changes. The reason for this is $\theta$ is only associated with the detector A and B, not C.

A value which is not used previously in the plots was used for basic test of the success of the method. For 1 MeV gamma-rays and A/T=0,34, B/T=0,32 and C/T=0,34 ratios, by simultaneous investigation of each three plots (Figs.3), the direction of the gamma-rays can be determined by about $\pm 10^o$ deviations from true values of $\theta = 45^o$ and $\phi = 60^o$.

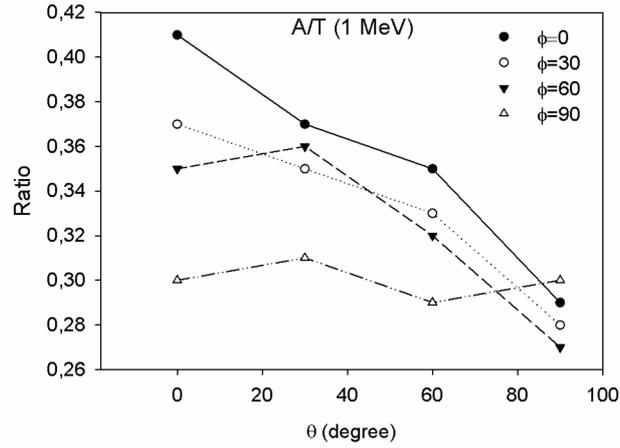

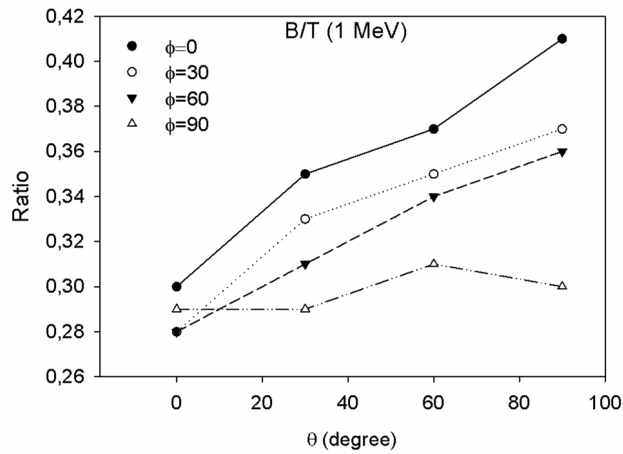

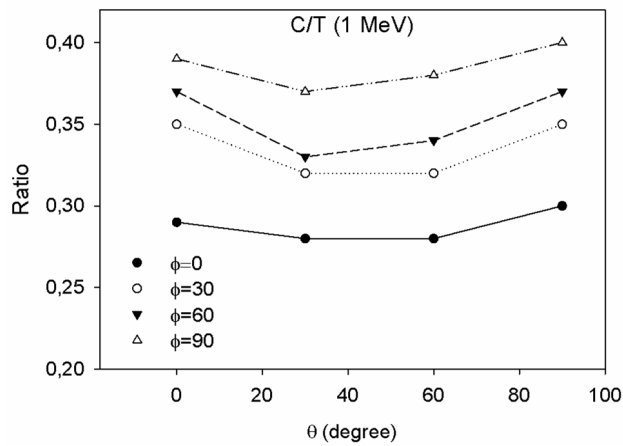

**Fig.3** The changing of detector photopeak count ratios by θ for 1 A/T (a), B/T (b) and C/T (c).

## 5. Conclusions

We have performed Geant simulations in order to demonstrate the ability of our designed detection system for identification of the gamma-ray directions. The method is based on the relative counts registered in different detectors in the array. According to the results, directional information of the gamma-rays was obtained by using developed method with $10^{o}$ accuracy. Besides the application of the method in space, this array can also be useful for measurements of terrestrial gamma-rays, determination of the location of radioactive sources and explanation dose rate differences found by car-borne measurement

**References**

[1] G.J. Fishman and C.A. Meegan, Annual Review of Astronomy and Astrophysics 33 (1995) 415-458.

[2] S.E. Woosley and J.S. Bloom,. 2006, arXiv:astro-ph/0609142v1.

[3] L.J. Gou et al., A&A 368 (2001) 464-470.

[4] B. Zhang, Advances in Space Research 40 (2007a) 1186-1198.

[5] D. W. Fox, et al., Nature 422 (2003) 284-286.

[6] Ph. Podsiadlowski et al., The Astrophysical Journal 607 (2004) L17-L20.

[7] T. Piran, Physics Reports 314 (1999) 575-667.

[8] J. van Paradijs, et al., Annual Review of Astronomy and Astrophysics 38 (2000) 379-425.

[9] P. Mészáros, Annual Review of Astronomy and Astrophysics 40 (2002) 137-169.

[10] B. Zhang and P. Mészáros, Inernational Journal of Modern Physics A 19 (2004) 2385.

[11] B. Zhang, Chinese Journal of Astronomy and Astrophysics 7 No.1 (2007b) 1-50.

[12] W.N. Johnson et al., The Astrophysical Journal Supplement Series 86 (1993) 693-712.

[13] R.M. Kippen et al., The Astropysical Journal 492 (1998) 246-262.

[14] D.J. Thompson, et al., Journal of Geophysical Research, 102 No.A7 PP.14 (1997) 735-14, 740.

[15] A. Gomboc, Contemporary Physics 53 No.4 (2012) 339-355.

[16] R. Aloisio, et al., Astroparticle Physics 29 (2208) 373-379.

[17] Y. Shirakawa, Nucl.Instrum.Meth. Phys. Res. B 213 (2004) 255-259.

[18] T. Tanimori, et al., 2004 arXiv:astro-ph/0403518v1.

[19] K. Fujimoto and Y. Noda, Radioactivity in the Environment 7 (2005) 118-125.

[20] Y. Shirakawa, Nucl.Instrum.Meth. Phys. Res. B 263 (2007) 58-62.

[21] E. Suarez-Garcia, et al., Nucl.Instrum.Meth. Phys. Res. A 624 (2010) 624-634.

[22] N. Schemm, et al., Proceedings of IEEE Sensors (2011) 1760-1763.

[23] G.J. Fishman, et al., Science 264 (1994) 1313-1316.

[24] S. Agostinelli, et al., Nucl.Instrum.Meth. Phys. Res. A 506 (2003) 250-303.